\documentclass[times,tighten]{aastex631}
\usepackage{amsmath}

\newcommand\cred{\color{red}}
\newcommand{\Msolar}{M_\odot}
\newcommand{\Mtot}{M_\text{tot}}

\newcommand{\Mg}{M_\text{gas}}

\newcommand{\ms}{m_*}
\newcommand{\mc}{m_\text{c}}
\newcommand{\mlow}{M_\text{l}}
\newcommand{\mup}{M_\text{u}}
\newcommand{\dotmac}{\dot{m}_\text{ac}^0}
\newcommand{\Gyr}{\text{Gyr}^{-1}}
\newcommand{\Msyr}{M_\odot\,\text{yr}^{-1}}

\def\ihep{Key Laboratory for Particle Astrophysics, Institute of High Energy Physics, Chinese Academy of Sciences, 19B Yuquan Road, Beijing 100049, P. R. China, wangjm@ihep.ac.cn}

\def\UCASast{School of Astronomy and Space Science, University of Chinese Academy of Sciences, 19A Yuquan Road, Beijing 100049, P. R. China}

\def\naocOptical{National Astronomical Observatories, Chinese Academy of Sciences (CAS), Beijing 100101, P. R. China}

\begin{document}


\title{\large\bf Chemical evolution of bulges of active galactic nuclei in the early Universe: 
roles of accreting stars}


\author[0009-0005-4152-2088]{Shuo Zhai}
\affiliation{\naocOptical}

\author[0000-0001-9449-9268]{Jian-Min Wang}
\affiliation{\naocOptical}
\affiliation{\ihep}
\affiliation{\UCASast}

\author[0000-0001-5841-9179]{Yan-Rong Li}
\affiliation{\ihep}

\author[0000-0001-9457-0589]{Wei-jian Guo}
\affiliation{\naocOptical}

\author[0000-0002-8980-945X]{Gang Zhao}
\affiliation{\naocOptical}
\affiliation{\UCASast}

\begin{abstract}

JWST/NIRCam observations reveal dense stellar cores in high-redshift galactic bulges, indicative of sustained star formation and potential stellar accretion.
We introduce accretion-modified star (AMS) as a new component in the chemical evolution of high-redshift bulges hosting active galactic nuclei (AGNs). 
The gas-phase chemical evolution of bulge environments containing AMS is modeled within 1 Gyr by combining population evolution and galactic chemical evolution formalisms, and observational signatures are tracked via photoionization modeling on Baldwin–Phillips–Terlevich (BPT) diagrams.
Sustained high accretion onto AMSs leads to rapid gas-phase metal enrichment of the bulge, producing abundance peaks up to five times solar metallicity within 0.1 Gyr and significantly modifying elemental ratios in the gas phase. 
Atypical gas-phase abundance patterns during early, high-accretion phases and gradually diminish as the accretion rate declines. 
In BPT diagrams, high-AMS-accretion scenarios shift the modeled emission-line sequence toward the local AGN branch and extend into the high-metallicity regime. 
Super-solar narrow-line regions observed in AGNs at $z \gtrsim 15$ may reflect such AMS-driven gas-phase enrichment of host bulge under extreme gas densities. 
While direct detection of AMSs within AGN bulges remains challenging, the model provides testable predictions for future spectroscopic surveys and motivates further exploration of non-canonical stellar populations in AGN host bulges.

\end{abstract}


\section{Introduction} \label{sec:intro}

The chemical abundance and its evolution in galactic bulges provide key insights into early star formation, black hole–bulge coevolution, and the buildup of galactic central regions \citep{Matteucci1990,Kormendy2004,Kobayashi2006,Genzel2010,Kormendy2013,Heckman2014,Barbuy2018}.
Galactic chemical evolution (GCE) models have evolved from early closed-box prescriptions \citep{Talbot1971} to frameworks incorporating gas flows, metal-dependent star formation, and stellar feedback \citep{Chiosi1980,Portinari2000,Matteucci2012,Matteucci2021}.
The key factors regulating chemical enrichment include the initial gas conditions\footnote{The initial conditions involve determining whether: the chemical composition of the initial gas is primordial or pre-enriched by a pre-galactic stellar generation; and the studied system is a closed box or an open system (infall and/or outflow).}, the stellar birthrate function $B(m_*,t)$\footnote{The birthrate function consists of two components: star formation rate (SFR) \citep{Schmidt1959} and initial mass function (IMF) \citep{Salpeter1955, Kroupa2001}, determining the total gas mass converted into stellar mass and the distribution of stellar masses.}, the stellar yields, and gas flows \citep{Matteucci2012, Matteucci2021}.
While GCE models account well for the metallicity distribution and abundance ratios in the stars of the galactic bulge \citep{Matteucci1990,Kobayashi2006,Ballero2007,Molero2024}, they are generally limited to standard star formation and feedback prescriptions, overlooking the potential contribution of stellar populations formed in high-density environments.


Recent observations and simulations indicate that bulge regions in high-redshift galaxies exhibit systematically higher gas densities compared to local counterparts. Cosmological simulations show that gas-rich, turbulent disks at early epochs fragment into massive clumps that migrate inward and coalesce into compact, dense bulges \citep{Ceverino2010, Bournaud2011}.
JWST/NIRCam imaging confirms that massive star-forming galaxies at $z \sim 2$ already contain dense central stellar cores, with near-infrared surface densities in the central kpc up to 7 times higher than those measured in optical bands \citep{Benton2024}.
Additional JWST observations further report the widespread presence of compact bulge-like structures in star-forming galaxies at $z \sim 2$–3, with central surface mass densities significantly exceeding those of local galaxies \citep{Chen2022,Finkelstein2022}.
At even earlier epochs ($z \gtrsim 6$), ultra-compact protobulge candidates have been identified, showing half-light radii below 100 pc and stellar mass densities comparable to those of massive local elliptical galaxies \citep{Carniani2024,Silk2024,Baker2025}.
High-redshift bulge environments therefore offer physical conditions including high density, efficient cooling, and strong pressure, which are favorable for the in-situ formation of accretion-modified star (AMS).


AMS was originally proposed in the environments of AGN accretion disks, where stars embedded in dense gas are expected to accrete mass efficiently and evolve rapidly \citep{Artymowicz1993, Cheng1999, Collin1999, Goodman2004, Wang2023}.
Recent simulations using Modules for Experiments in Stellar Astrophysics (MESA) suggest that accreting stars can grow into very massive objects exceeding $100\,\Msolar$ within short timescales \citep{Cantiello2021, Jermyn2021, Jermyn2022, Ali2023,Fryer2025}.
Following core collapse, these stars are expected to leave behind compact remnants and produce helium and heavy elements that pollute the AGN disk.
Given the similarly high gas densities in AGN disks and high-redshift bulge environments, AMS can extend to bulge regions as a potential in-situ stellar population.

Motivated by the above considerations, we develop a chemical evolution model that incorporates AMS as sources of rapid gas-phase enrichment and investigate their impact on bulge chemistry. By varying the AMS fraction, the model tracks the evolution of $\alpha$-elements and Fe-peak elements in the bulge gas. The resulting gas abundance patterns are then input into the MAPPINGS photoionization code \citep{Sutherland2018} to simulate emission-line signatures on the Baldwin–Phillips–Terlevich (BPT) diagram \citep{Baldwin1981}, enabling an assessment of whether AMS can leave spectroscopic imprints in AGN bulge environments.

The paper is organized as follows. We describe the chemical evolution model with AMS and photoionization model grids in Section~\ref{sec:CEM} and Section~\ref{sec:Photoionization}, respectively. The chemical evolution results and photoionization results are presented in Section~\ref{sec:GCEresult} and Section~\ref{sec:VOresult}, respectively. Several model caveats and observational implications are discussed in Section~\ref{sec:modeldependences}. Finally, our conclusions are listed in Section~\ref{sec:concl}. Throughout this work, we adopt a flat $\Lambda$CDM cosmology with $H_0=67.4\, \mathrm{km\, s^{-1}\, Mpc^{-1}}$, $\Omega_\Lambda = 0.7$, and $\Omega_m =0.3$ \citep{Planck2020}.

\section{Chemical Evolution Model with Accretion-Modified Stars} \label{sec:CEM}


Motivated by the high-density conditions revealed by simulations and JWST observations, we construct a closed-box chemical evolution model for AGN bulges (within 1 Gyr, $z \sim 5.7$) that explicitly incorporates the contribution of AMS. 
The gas-phase evolution is modeled by combining stellar population formalisms with GCE formalisms. The physical parameters of the bulge, stellar, and AMS are summarized in Table~\ref{tab:GlobalModel}.
Section~\ref{subsec:GoverningEq}–\ref{subsec:Yields} detail the model components, including governing equations for mass function and elemental abundance evolution ($\S$~\ref{subsec:GoverningEq}), the star formation law ($\S$~\ref{subsec:SFR}), the IMF ($\S$~\ref{subsec:IMF}), the parameterization of AMS accretion and environmental suppression ($\S$~\ref{subsec:AMS}), and stellar yields and lifetimes ($\S$~\ref{subsec:Yields}).

\subsection{Governing Equations of Mass Functions and Chemical Abundances} \label{subsec:GoverningEq}

Following the GCE model, we adopt a simple closed-box framework as a baseline to highlight how stellar-scale accretion reshapes bulge chemical enrichment, while infall and outflow are not considered. The gas within the bulge is assumed to follow instantaneous mixing. The initial state of the bulge is assumed to be a homogeneous sphere of primordial gas (75\%\,H and 25\%\,He) with a total mass of $10^{10}\,\Msolar$. Galactic winds are not considered, as the potential well of the bulge is excessively deep for self-regulation.

Based on the above assumptions, we model the evolution of the AMS mass function under the combined effects of star formation, accretion, and supernova explosion.
The continuity equation in mass space, as formulated by \citet{Wang2023}:
%
\begin{gather}
    \frac{\partial}{\partial t} \Psi (m_*,t) +
    \frac{\partial}{\partial m_*} [\dot{m}_* \Psi (m_*,t)] =
    \psi(t) \phi(m_*)
    -\dot{\mathcal{R}}_\mathrm{SN} (m_*,t) , \\
    \dot{\mathcal{R}}_\mathrm{SN} (m_*, t) \approx 
    \begin{cases}
    {\displaystyle{\frac{\Psi (m_*, t)}{\tau_\mathrm{evo} (m_*)},}}& t \geq \tau_\mathrm{evo}, \\
    0 ,& t < \tau_\mathrm{evo}, 
    \end{cases}
	\label{eq:MFsingle}
\end{gather}
where $\dot{m}_*$ is the accretion rate of stars, $\psi(t)$ is the star formation rate (SFR), $\phi(m_*)$ is the IMF, $\dot{\mathcal{R}}_\mathrm{SN}$ is supernova rate, and $\tau_\mathrm{evo}$ is the lifetime of stellar main-sequence. 

In conjunction with the general GCE formalism \citep{Matteucci2012}, the time derivative of the gas-phase mass of a chemical species can be expressed as
%
\begin{equation}
    M_{\rm gas}(t) = \sum_{\rm X} M_{\rm X,gas}(t),
\end{equation}
\begin{equation}
    \dot{M}_{\rm X,gas}(t) = -\frac{M_{\rm X,gas}}{M_{\rm gas}} \psi(t) + \int_{m_{\rm l}}^{m_{\rm u}} m_* \frac{\partial}{\partial m_*} [\dot{m}_* \Psi (m_*,t)] dm_*
    + \int_{m_{\rm l}}^{m_{\rm u}} \dot{\cal R}_{\rm SN}(m_*,t)\, Y_{\text{X},m_*} dm_*,
    \label{eq:Mgasi}
\end{equation}
where $Y_{\text{X},m_*}$ represents the yield of element X from a star with mass $m_*$, $M_{\rm X,gas}$ is the gas-phase mass of the element X, and $M_{\rm gas}$ is the total gas-phase mass of the bulge.
The abundance of the chemical elements relative to solar abundance can be expressed as
\begin{equation}
    \text{[X/H]}(t) = 
    \text{log}_{10} \left ( \frac{M_\text{X,gas}(t)}{\mathcal{A}_\text{X} M_\text{H}(t)} \right )
    - \text{log}_{10} \left ( \frac{N_\text{X}}{N_\text{H}} \right )_\odot ,
	\label{eq:metal}
\end{equation}
where $\mathcal{A}_\text{X}$ is the mass number of chemical species X ($\mathcal{A}_\text{H}=1$, $\mathcal{A}_\text{He}=4$, $\mathcal{A}_\text{C}=12$, $\mathcal{A}_\text{N}=14$, $\mathcal{A}_\text{O}=16$, $\mathcal{A}_\text{Ne}=20$, $\mathcal{A}_\text{Mg}=24$, $\mathcal{A}_\text{Si}=28$, $\mathcal{A}_\text{S}=32$, $\mathcal{A}_\text{Ca}=40$, $\mathcal{A}_\text{Fe}=56$), and $(N_\text{X}/N_\text{H})_\odot$ is the solar abundance. In this paper, we adopt 
the solar abundance from \citet{Asplund2009} with $Z_{\odot}=0.0134$.

\subsection{The Star Formation Rate} \label{subsec:SFR}

For spheroids, such as ellipticals and bulges, the most common prescription for SFR is
\begin{equation}
    \psi(t) = \nu \left [ \frac{M_\mathrm{gas}(t)}{M_\mathrm{tot}} \right ]^k ,
        \label{eq:SFR}
\end{equation}
where $k$ is a dimensionless parameter depending on the morphology, $\nu$ is the star formation efficiency (SFE), $\Mg$(t) is the gas-phase mass in the bulge at time $t$, and $\Mtot$ is the total baryonic mass of the bulge \citep{Kennicutt1998,Matteucci2012}. We assume $k=1$ and $\nu = 20~\Gyr$ \citep{Ballero2007}.

\subsection{The Initial Mass Function} \label{subsec:IMF}

The IMF is assumed to be constant in time and space for conventional GCE and is generally expressed as a power law
\begin{equation}
    \phi(m_*) = \phi_0 m_*^{-\alpha} ,
	\label{eq:IMF}
\end{equation}
where $\phi_0$ is a constant derived from the normalization of the mass-weighted IMF
\begin{equation}
    \int_{M_\text{l}}^{M_\text{u}} m_* \phi(m_*)dm_* = 1.
	\label{eq:norIMF}
\end{equation}
We consider a Salpeter IMF with $\alpha=2.35$ \citep{Salpeter1955}.
The IMF prescription serves as the initial distribution onto which AMS-driven mass growth is applied in the next section.

The mass range of the input is $\mlow-\mup=0.1-100\,\Msolar$, which is a conventional choice in GCE studies \citep{Ballero2008,Matteucci2012,Matteucci2021} and is consistent with the process of star formation in normal molecular clouds \citep{McKee2007}. 
The mass range differs significantly from the stellar mass range adopted for AGN disks ($1.0-10\,\Msolar$) in \citet{Wang2023}, due to the distinct star formation processes in highly dense environments such as AGN disks compared to normal molecular clouds.

\subsection{Accretion-Modified Star} \label{subsec:AMS}

AMS refers to star that grow in mass by accreting gas from their surroundings, particularly in dense environments such as AGN disks \citep{Artymowicz1993, Cheng1999, Collin1999, Goodman2004}.
Recent studies using MESA have shown that AMS in AGN disk can grow into very massive stars exceeding $100\,\Msolar$ and undergo rapid evolution \citep{Cantiello2021, Jermyn2021, Jermyn2022, Ali2023, Fryer2025}.

Our model considers AMS formation and evolution in bulge environments, where stars form in situ and subsequently grow via cold gas accretion. After core collapse, AMSs leave behind compact remnants and eject processed material into the interstellar medium, thereby altering the chemical composition of the host bulge.
Since the accretion process remains poorly understood and has previously been treated using phenomenological models, we adopt a similar approach by modifying the prescription of \citet{Wang2023} and include gas depletion effects from a closed-box model:

%
\begin{equation}
    \dot{m}_*(t) = 
    \dot{m}_\text{ac}^0
    \left[ \frac{M_\text{gas}(t)}{M_\text{tot}} \right]
    \left ( \frac{m_*}{M_\odot} \right )^2
    \left ( 1 + \frac{m_*}{m_\text{c}} \right )^{-\beta} ,
	\label{eq:AMS}
\end{equation}
where $M_{\rm gas}(t)/M_{\rm tot}$ accounts for gas depletion in the closed-box scenario, $\dotmac$ is a characteristic parameter depending on the mean particle density of the bulge, $\mc$ is the critical mass above which accretion is suppressed, and $\beta$ controls the strength of the suppression, capturing environmental effects, primarily turbulence-induced suppression.

To constrain the plausible range of AMS accretion rates in bulge environments, we estimate an upper limit based on the mean gas density profile of the bulge. 
The characteristic accretion rate $\dotmac$ can be estimated using the classical Hoyle–Lyttleton–Bondi prescription as:
\begin{equation}
    \dotmac \simeq 3.5 \times 10^{-7}\, M_\odot\,\mathrm{yr^{-1}}
    \left( \frac{\bar{n}}{6 \times 10^{4}\,\mathrm{cm^{-3}}} \right)
    \left( \frac{m_*}{1\,M_\odot} \right)^{2}
    \left( \frac{v}{1\,\mathrm{km\,s^{-1}}} \right)^{-3},
    \label{eq:mac0}
\end{equation}
where $\bar{n}$ is the mean protobulge particle density, and $v$ represents the effective relative velocity between the star and the surrounding gas, incorporating both thermal and turbulent components. 
We adopt $\bar{n}\approx 6\times 10^4 \, M_{10} \, r_{150}^{-3} \, \rm cm^{-3}$, where $r_{150}=r_{\rm eff}/150\,\rm pc$ is the median observed radius for a galaxy with mass $M_{10}=M_{\rm tot}/10^{10}\,\Msolar$ \citep{Silk2024}.
Therefore, the upper limit is estimated to be $\dotmac \approx 10^{-7}\,\Msyr$ for protobulge environments.
Note that this estimate does not account for the radial distribution of the bulge, thus leading to a certain degree of overestimation. Based on this estimate, we adopt fiducial values of $\dotmac = 10^{-8} \, \mathrm{and}\, 10^{-7}\,\Msyr$ to represent low-$\dotmac$ and high-$\dotmac$ regimes, respectively. At lower densities, for example, $\dotmac = 10^{-9}\,\Msyr$, AMS growth is negligible, as stars typically complete their lifetimes before accreting significantly. In contrast to AGN disks, where densities are orders of magnitude higher ($\rho_{\rm AD}\sim10^{-12}\,\rm g\,cm^{-3}$), AMSs in bulges evolve under a lower gas supply. Consequently, the mass functions of AMS in bulges tend to remain closer to the input IMF and struggle to develop top-heavy features, even at high accretion rates.

The parameter $\beta$ serves as a simplified representation of the suppression of stellar accretion due to environmental factors, primarily turbulence driven by stellar feedback and large-scale inflows \citep[e.g.,][]{Krumholz2006,Bacchini2020,Forbes2023}.
These processes are known to reduce accretion efficiency compared to the idealized Bondi-Hoyle prediction, but their precise impact is difficult to quantify in global chemical evolution models \citep{Burleigh2017}. We therefore adopt a constant value of $\beta$ in the range $1.0-2.0$  \citep{Toyouchi2022, Wang2023}, which broadly captures the plausible strength of environmental regulation.

\subsection{Stellar Lifetimes and Yields} \label{subsec:Yields}

Since stellar lifetime approximates its main-sequence lifetime, we employ the polynomial fitting result of Population III main-sequence lifetime$-$mass relation from \citet{Schaerer2002}:
\begin{equation}
    \text{log} \left ( \frac{\tau_\text{evo}}{\text{Gyr}} \right ) = 
    1.01 - 4.43x + 2.09x^2 -0.47x^3 +0.04x^4,
	\label{eq:stellarlifetime}
\end{equation}
where $x=\text{log}_{10}(m_*/M_\odot)$. As shown in Appendix \ref{app:lifetime}, 
the lifetime of main-sequence stars spans from Myr to Gyr for stellar masses ranging from $0.1-1000.0\,\Msolar$, and is insensitive to metallicity.

The time-dependence of the abundances of hydrogen, helium, carbon, nitrogen, oxygen, neon, magnesium, silicon, sulfur, calcium, and iron is calculated using the following stellar yield models. For low-to-intermediate-mass stars (LIMs), we apply the yields of \citet{Karakas2010}, covering a range in metallicity $Z = 0.0001, 0.004, 0.008 \,\rm and\,0.02$, and masses between $1\,\Msolar$ and $6\,\Msolar$, supplemented by \citet{Cristallo2015}. For massive stars, the core-collapse supernovae (CCSNe) yields are taken from the R set\footnote{Set R, tab\_yieldstot\_iso\_exp.dec, assuming mixing and fall-back, and the mass cut is chosen so that each supernova event ejects 0.07 $\Msolar$ of $^{56}\text{Ni}$.} of \citet{Limongi2018} with zero initial rotational velocities (IRV = 0), covering four metallicities (i.e., [Fe/H] = 0, -1, -2, and -3) and masses range from $13\,\Msolar$ to $120\,\Msolar$. For very massive stars, we adopt the yields of \citet{Portinari1998}, in a range of masses from $120\,\Msolar$ to $1000\,\Msolar$ and metallicities Z = 0.0004, 0.004, 0.008, 0.02, 0.05. 
The yields corresponding to the upper limit of the known metallicity are utilized without extrapolation if the metallicity exceeds the established range, despite the potential for overestimating the gas-phase metals of bulges. Fortunately, subsequent calculations indicate that the state of $Z\textgreater0.05$ is short-lived, thus justifying the approximation.

\begin{table*}
	\centering
	\caption{Parameters of the Global Model of the bulge with AMS}
	\label{tab:GlobalModel}
	\begin{tabular}{llc}
	\hline
        \hline
	Parameter                  & Description                                                                                   & Values    \\
	\hline
        Bulge physics & & \\
        \hline
        $\Mtot$ ($\Msolar$)         & Total baryonic mass of the bulge                                                              & $10^{10}$ \\
        $R$ (kpc)                   & Radius of the bulge                                                                           & 1.0       \\
        \hline
        Stellar physics & & \\
        \hline
        $\alpha$                    & Index of the IMF                                                                              & 2.35\\
        $\mlow,\,\mup$ ($\Msolar$)  & Lower and upper mass limits of IMF                                                            & 0.1, 100  \\
        $\nu$ ($\text{Gyr}^{-1}$)   & Star formation efficiency                                                                     & 20     \\
        \hline
        AMS physics & & \\
        \hline
        $\mc$ ($\Msolar$)           & The critical mass of stars with depressed accretion in the context of radiation environments    & 10        \\
        $\dotmac$ ($\Msyr$)         & Accretion rate of $1\,\Msolar\,$AMS                                                           & $10^{-8}$, $10^{-7}$ \\
	$\beta$                     & Index of correction of AMS accretion rates due to environment                                 & 1.0, 1.5, 2.0 \\
        \hline
	\end{tabular}
\end{table*}

\section{Photoionization Modeling with Chemically Evolved Bulge Abundances} \label{sec:Photoionization}

Given that directly identifying AMS in high-redshift bulges remains observationally challenging, we adopt a theoretical approach to evaluate whether AMS-driven chemical enrichment produces observable signatures. 
To this end, we employ the photoionization code MAPPINGS V (v5.2.0, \citealt{Sutherland2018}) to compute theoretical line ratios on standard optical diagnostic diagrams.
Specifically, we use the chemical abundances predicted by our bulge evolution model (Section \ref{sec:CEM}) as input to MAPPINGS and track the evolution of the [\ion{N}{2}]/H$\alpha$ and [\ion{O}{3}]/H$\beta$ line ratios on the BPT diagram.

In this framework, AMSs are treated solely as enrichment sources, while AGNs provide the dominant radiation field for photoionization. 
We adopted the AGN radiation field as the ionizing source since AGNs at high redshift are among the most luminous and detectable objects. Here we neglect the contribution of AMS radiation because our models are designed to explore luminous AGN-dominated systems that are more readily observable at high-redshift. Quantitatively, even for a high SFR of $\psi \sim 10 - 30\,M_\odot\,\mathrm{yr}^{-1}$, the UV luminosity of the AMS population at 1500\,\AA\ is $\nu L_\nu \lesssim 10^{44}\,\mathrm{erg\,s^{-1}}$ \citep{Kennicutt1998ARA}. In contrast, adopting the bolometric correction from \citep{Richards2006}, the corresponding 1500\,\AA\ luminosity for AGNs with $10^{7-8}\,M_\odot$ supermassive black holes radiating at the Eddington limit is $(0.6-6)\times10^{45}\,\mathrm{erg\,s^{-1}}$, confirming that the ionizing contribution from AMSs is subdominant in luminous systems.
Furthermore, recent JWST observations of $z\textgreater6$ quasars show that most sources exhibit near- or super-Eddington accretion \citep{Yang2023}, and many lie above the local black hole–host mass relation, consistent with heavy seed and/or super-Eddington scenarios \citep{Harikane2023, Maiolino2023, Marshall2023, Ubler2023, Kocevski2023}.
We therefore select the OPTXAGNF radiation field model \citep{Done2012,Jin2012,Jin2017} with a high Eddington ratio as the ionizing continuum.


We adopt a differential approach: the outputs from models with and without AMSs are separately fed into MAPPINGS, allowing a direct comparison of the predicted photoionization tracks. The gas-phase abundance is treated as the only variable, whereas all other input parameters (e.g., radiation fields, dust properties, and density structure) are held fixed to isolate the impact of AMS-driven enrichment. This approach allows us to use the BPT evolution trajectory as a diagnostic tool to probe the potential presence of AMSs in high-redshift bulges.




Finally, the photoionization grid is created with the following configurations:
\begin{itemize}
  \item[1.]A plane parallel geometry.
  \item[2.]Ionization parameter at the inner edge set in the range $-3.5 \leq \mathrm{log}\, U \leq -1.0$, uniformly spaced at a logarithmic interval of 0.5\,dex.
  \item[3.]Initial gas pressure $\mathrm{log}\, P/k\, (\mathrm{cm^{-3}\,K}) = 7.0$.
  \item[4.]OPTXAGNF radiation fields \citep{Done2012,Jin2012,Jin2017} with high Eddington ratio.
  \item[5.]Chemical composition using the results of the chemical evolution in Section \ref{sec:GCEresult} (with and without AMSs), sampling at a logarithmic interval of 0.2\,dex in the range of $10^7-10^9\,$yr. For the elements not included in our calculations, tests in Appendix~\ref{app:testAbn} have indicated their minimal impacts on the results, and so their abundances are assumed to be extremely low (logarithmically -10) and constant.
  \item[6.]A set of depletion factors onto dust grains based on iron being 96.8\% depleted ($\mathrm{log}(\mathrm{Fe_{free}}/\mathrm{Fe_{total}})=-1.5$; \citealt{Jenkins2009, Thomas2018}).
  
  \item[7.]No dust destruction.
\end{itemize}

\section{Chemical Evolution Results with Accretion-Modified Stars} \label{sec:GCEresult}


This section presents the chemical evolution results of AMSs within 1 Gyr ($z \sim 5.7$) for a bulge with a total mass of $10^{10}\,M_\odot$, assuming an IMF slope of $\alpha = 2.35$ and a SFE of $\nu = 20\,\mathrm{Gyr^{-1}}$. 
The impact of varying AMS accretion parameters, with $\dot{m}_{\mathrm{ac}}^0$ ranging from $10^{-8}$ to $10^{-7}\,M_\odot\,\mathrm{yr^{-1}}$ and $\beta$ from 1.0 to 2.0, is evaluated in terms of the stellar mass function, supernova rates, gas-phase metallicity, and elemental abundances.
A comparison between models with and without AMSs is also conducted to assess their cumulative impact on bulge evolution.

\subsection{Stellar Mass Function Evolution} \label{subsec:AMSMF}

\begin{figure*}
    \includegraphics[width=\textwidth]{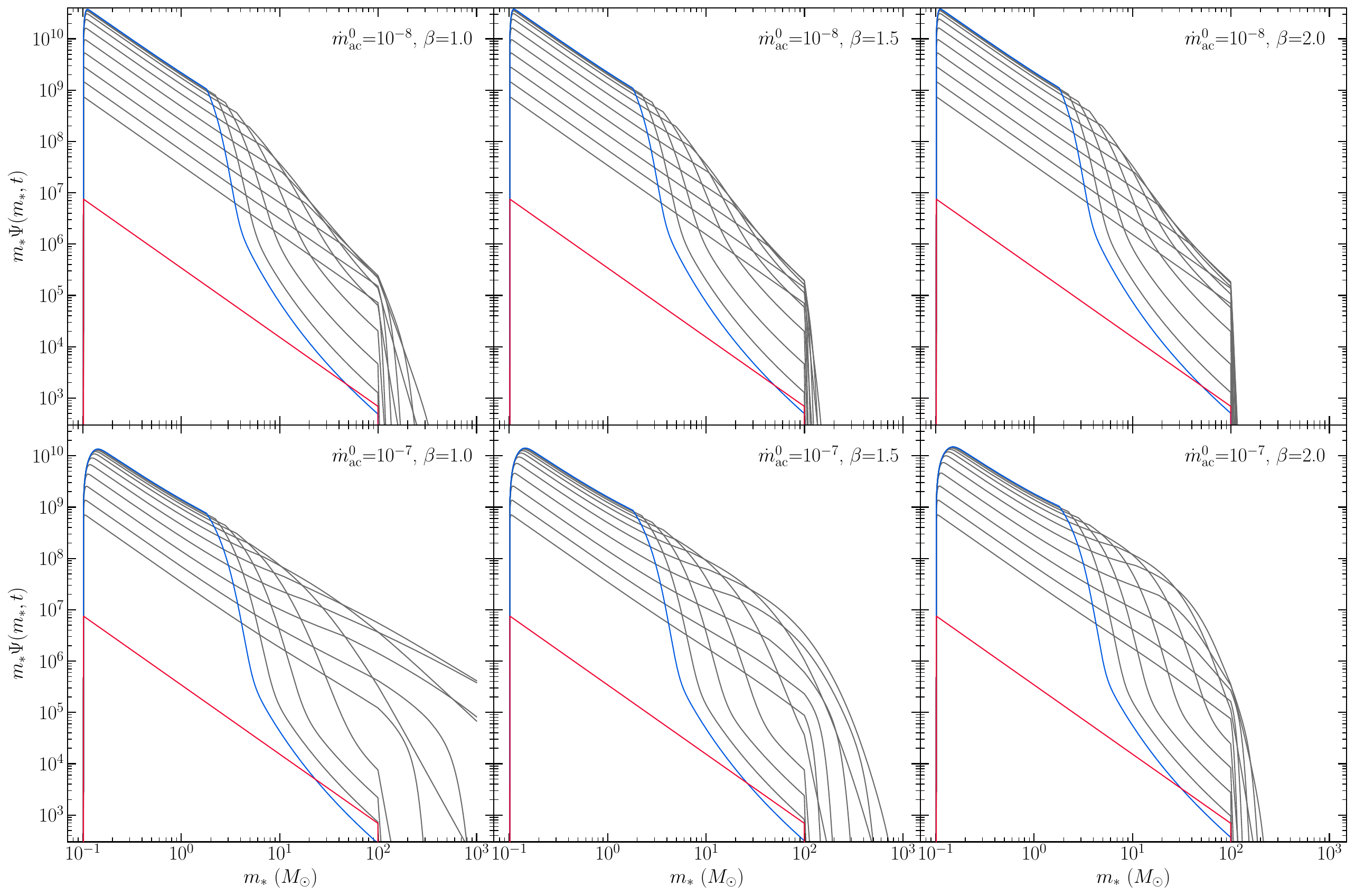}
    \caption{\footnotesize
    The evolving AMS mass functions for a bulge with a total mass of $10^{10}\,\Msolar$, an SFE of 20 $\Gyr$, and a Salpeter ($\alpha$=2.35) IMF. The panels are arranged as follows: horizontal rows from left to right with increases of $\beta$ for a given $\dotmac$, and vertical with increases of $\dotmac$ from upper to below for a given $\beta$. The red and blue curves represent the 
    initial and final (1\,Gyr) stages of the evolution, respectively. The gray curves are uniformly distributed at a logarithmic interval of 0.3\,dex in the range of $10^6-10^9\,$yr.}
    \label{fig:MF_2.35}
\end{figure*}

The evolution of AMS mass functions under varying $\dot{m}^0_{\mathrm{ac}}$ and $\beta$ is illustrated in Figure~\ref{fig:MF_2.35}. We first fix $\beta$ and explore the dependence on $\dot{m}^0_{\mathrm{ac}}$. At low $\dot{m}^0_{\mathrm{ac}}$, mass functions evolve slowly, while increasing $\dot{m}^0_{\mathrm{ac}}$ leads to the formation of a significant population of very massive stars. For instance, a 1 $M_\odot$ AMS can reach 10 $M_\odot$ within $\sim 10^7$ yr under the condition of $(\dot{m}^0_{\mathrm{ac}}, \beta) = (10^{-7}, 1.0)$.
We then fix $\dot{m}^0_{\mathrm{ac}}$ and vary $\beta$ to examine the environmental suppression of accretion. 
Higher $\beta$ generally enhances the formation of top-heavy mass functions (as seen in AGN disks; \citealt{Wang2023}). 
However, we find that such a distribution does not emerge even under extreme suppression conditions $(\dot{m}^0_{\mathrm{ac}}, \beta) = (10^{-7}, 2.0)$, mainly due to the significantly lower gas density in bulges relative to AGN disks.
Furthermore, the AMS accretion rate in our model declines with decreasing gas density, further limiting the formation of very massive stars. As a result, the final shapes of the mass function are broadly similar in different combinations of $\dot{m}^0_{\mathrm{ac}}$ and $\beta$.

%

\subsection{Supernova Rate and Type Distribution} \label{subsec:SNrate}

AMSs alter the stellar distribution, which in turn impacts the evolution of supernova rates. Stars with masses above approximately $8\,\Msolar$ explode as CCSNe, producing mainly $\alpha$-elements \citep{Woosley2002,Nomoto2013}. LIMs primarily produce helium, carbon, and nitrogen during the asymptotic giant branch phase and end their lives as carbon-oxygen white dwarfs, typically without undergoing supernova explosions \citep{Nomoto2013}.

\begin{figure*}
    \includegraphics[width=\textwidth]{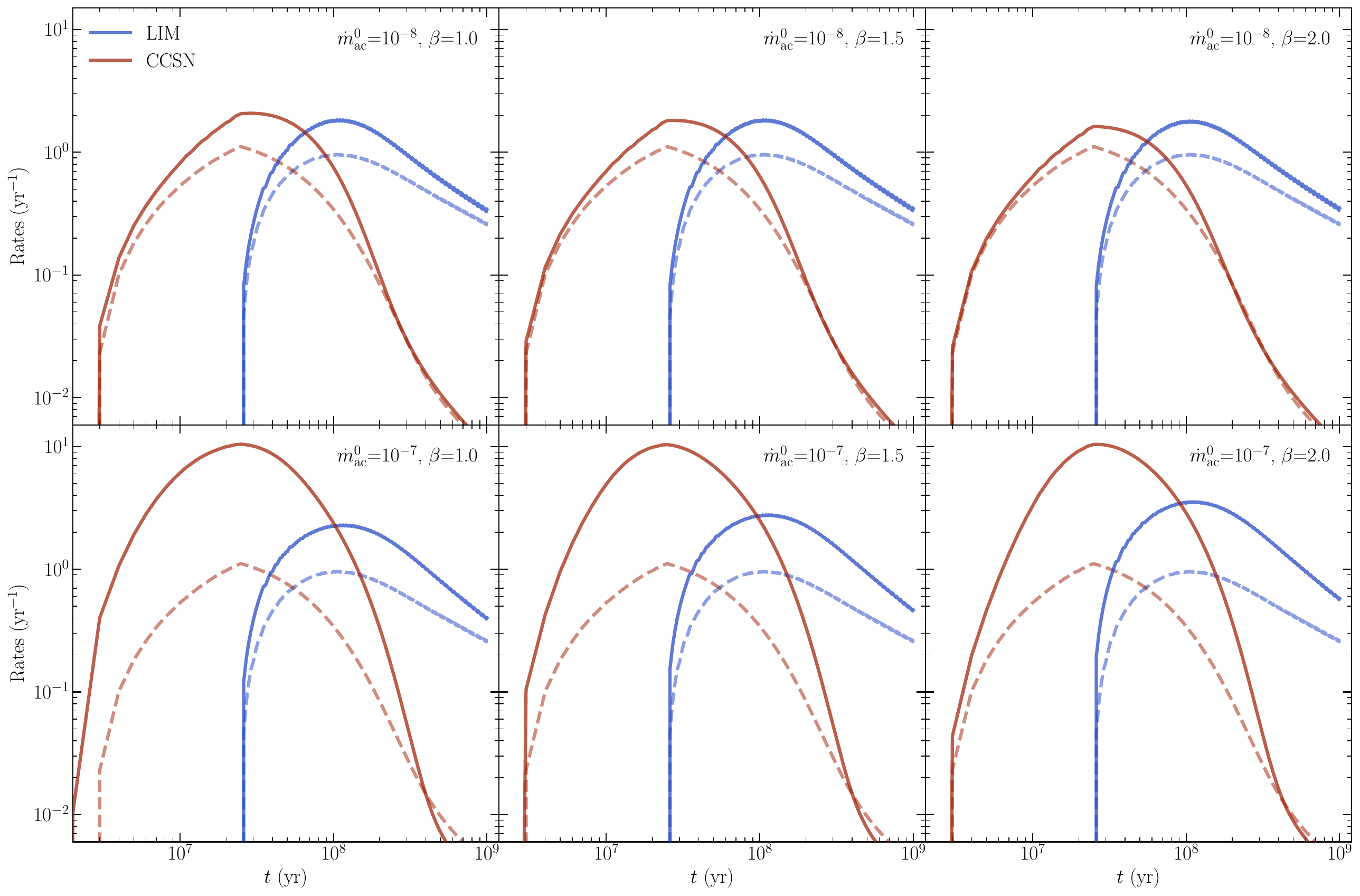}
    \caption{\footnotesize
    Evolution of the rates of type II supernova for a bulge with a total mass of $10^{10}\,\Msolar$, an SFE of 20 $\Gyr$, and a Salpeter ($\alpha$=2.35) IMF. The panels are arranged as follows: horizontal rows from left to right with increases of $\beta$ for a given $\dotmac$, and vertical with increases of $\dotmac$ from upper to below for a given $\beta$. The solid and dashed curves represent the AMS and conventional GCE scenarios, respectively. The blue and red curves represent LIMs and CCSNe, respectively.}
    \label{fig:SN_2.35}
\end{figure*}

Figure~\ref{fig:SN_2.35} shows the evolution of LIM and CCSN rates under varying AMS accretion parameters ($\dotmac$, $\beta$). 
At high-$\dotmac$, the CCSN rate increases by nearly an order of magnitude compared to the conventional case, and the LIM rate increases overall with a peak approximately twice as high.
Increasing $\beta$ leads to a moderate reduction in CCSN and a slight enhancement in LIM production, reflecting the role of $\beta$ in shifting the stellar mass distribution toward lower-mass progenitors. 
These trends indicate that both the timing and intensity of supernova explosion are sensitive to the AMS accretion parameters.
A high accretion rate combined with a low $\beta$ value can trigger earlier and more pronounced supernova activity, potentially altering the chemical enrichment history and star formation regulation in galactic bulges.


%

\subsection{Metallicity Evolution} \label{subsec:Metallicity}

\begin{figure*}
    \includegraphics[width=0.5\textwidth]{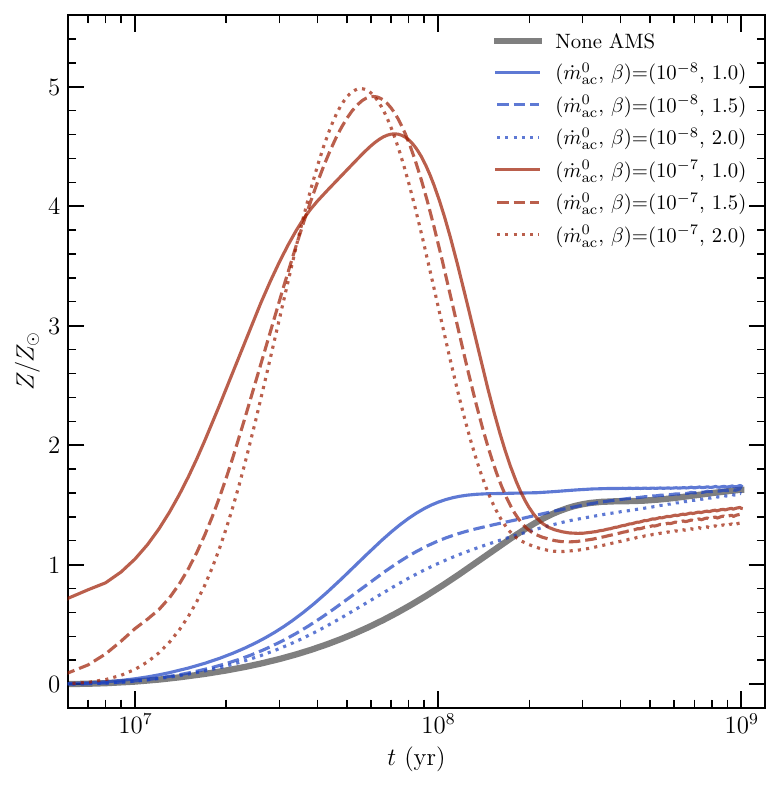}
    \caption{\footnotesize
    Time-dependence of metallicity for a bulge with a total mass of $10^{10}\,\Msolar$, an SFE of 20 $\Gyr$, and a Salpeter ($\alpha$=2.35) IMF. Gray, blue, and red curves denote conventional GCE scenario, AMS with low-$\dotmac$ and AMS with high-$\dotmac$, respectively. Solid, dashed, and dotted lines represent models with $\beta = 1.0$, 1.5, and 2.0, respectively.}
    \label{fig:Z_AMS_IMF_nu}
\end{figure*}

Figure~\ref{fig:Z_AMS_IMF_nu} illustrates the impact of AMS parameters $(\dotmac,\,\beta)$ on the gas-phase metallicity evolution. 
The overall trend reflects how AMSs accelerate and affect the chemical enrichment history of bulges.
At high-$\dotmac$, metallicity enrichment proceeds rapidly, reaching solar metallicity within approximately $10^7\,$yr, about an order of magnitude earlier than in the conventional scenario. The metallicity subsequently peaks at around $5\,Z_\odot$, and the super-solar phase can persist for up to $10^8$ yr, forming a prominent bump. 
As $\beta$ increases, the peak metallicity becomes higher and sharper, while the duration of the super-solar phase slightly shortens.
In contrast, at low-$\dotmac$, the enrichment track closely follows that of the conventional scenario. The effect of $\beta$ becomes more subtle in this regime, primarily acting to slightly delay or flatten the enrichment.
Despite differences in early-time behavior, all models eventually converge to similar metallicity levels after around $2 \times 10^8$ yr. 
The reason for the eventual convergence in metallicity, as well as the physical interpretation of the transient super-solar phase and its $\beta$-dependence, is discussed in Section~\ref{subsec:comparison}.


%

\subsection{Elemental Abundance Trends} \label{subsec:AMSAbun}

\begin{figure*}
    \includegraphics[width=\textwidth]{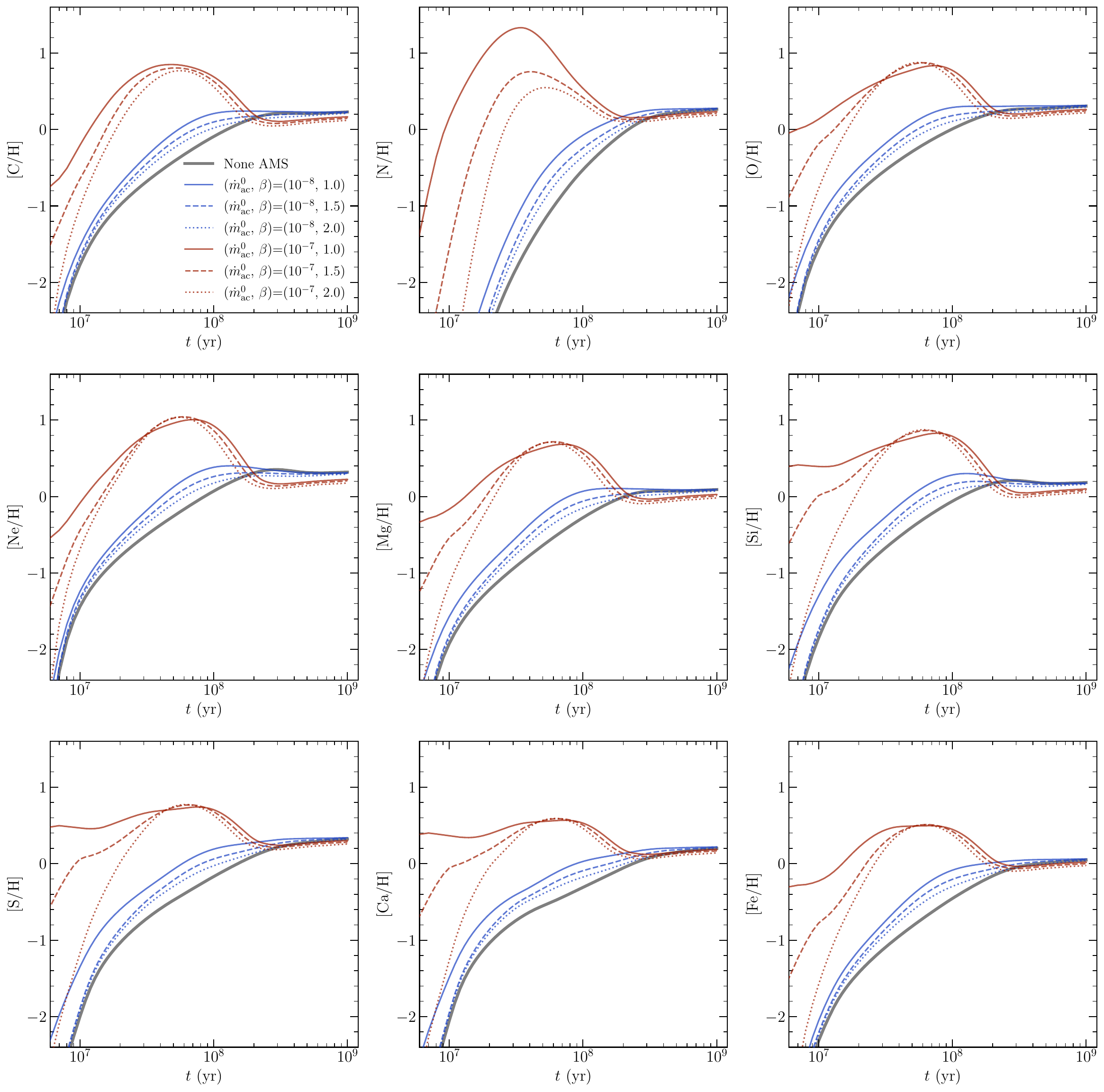}
    \caption{\footnotesize
    Time-dependence of [X/H] abundance ratios for $\alpha$-elements (O, Ne, Mg, Si, S, and Ca), C, N, and Fe in a bulge with a total mass of $10^{10}\,\Msolar$, an SFE of $20\,\Gyr$, and a Salpeter IMF. The panel layout same as Figure \ref{fig:MF_2.35}. The different curves denote different elements as indicated in the upper left panel. The abundance of each element in the solar is as follows: $\rm [C/H]_{\odot}=-3.57$, $\rm [N/H]_{\odot}=-4.17$, $\rm [O/H]_{\odot}=-3.31$, $\rm [Ne/H]_{\odot}=-4.07$, $\rm [Mg/H]_{\odot}=4.4$, $\rm [Si/H]_{\odot}=-4.49$, $\rm [S/H]_{\odot}=-4.88$, $\rm [Ca/H]_{\odot}=-5.66$, and $\rm [Fe/H]_{\odot}=-4.5$.}
    \label{fig:Metal_2.35}
\end{figure*}
\begin{figure*}
	\includegraphics[width=0.5\textwidth]{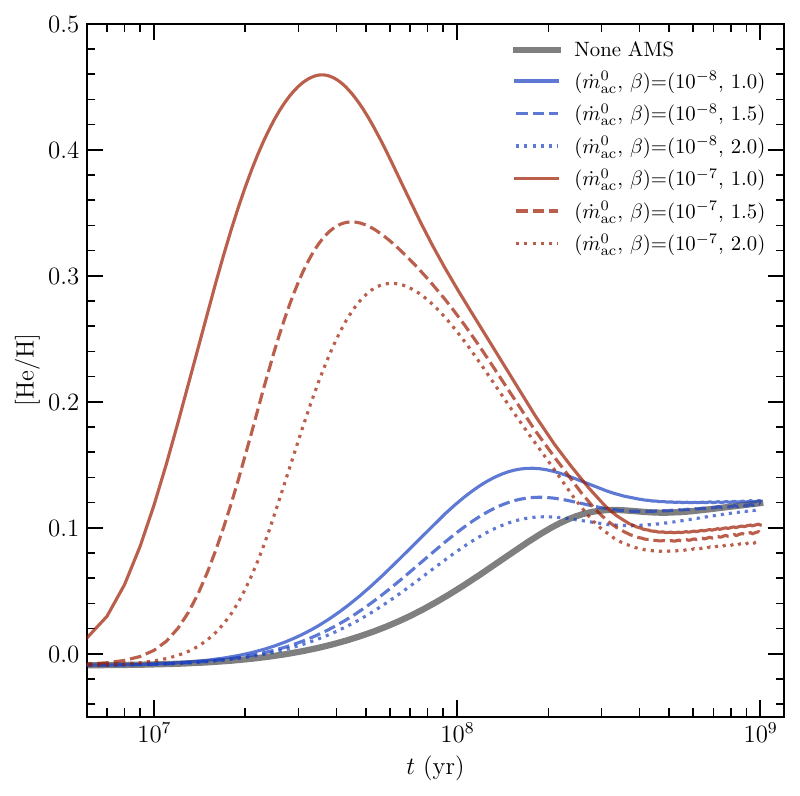}
    \caption{\footnotesize
    Time-dependence of [He/H] abundance ratios in a bulge with a total mass of $10^{10}\,\Msolar$, an SFE of $20\,\Gyr$, and a Salpeter IMF. The different curves denote different AMS parameters as indicated in the upper right panel. The He abundance in the solar is $\rm [He/H]_{\odot}=-1.07$.}
    \label{fig:He_2.35}
\end{figure*}

Figures~\ref{fig:Metal_2.35}-\ref{fig:He_2.35} show the evolution of elemental abundances for representative species, including $\alpha$-elements (O, Ne, Mg, Si, S, and Ca), C, N, Fe, and He. With AMSs included, the enrichment of all elements proceeds significantly faster than in the conventional chemical evolution scenario. At high-$\dotmac$, most elements reach solar abundance within $\sim 10^7$ yr and exhibit pronounced super-solar peaks. In contrast, models with low-$\dotmac$ show only modest deviations from the conventional case, with enrichment proceeding more gradually and peak values only slightly above solar.

At high-$\dotmac$, the prominent overabundances arise from the rapid formation of very massive stars via prolonged accretion, contributing large yields of alpha-elements, carbon, and iron via core-collapse or pair-instability supernovae.
Consequently, elemental abundances increase significantly during the first $1-5\times10^7\,$ yr.
The early-time enhancement is more pronounced for lower values of $\beta$, which delay the decay of accretion rates and facilitate the formation of additional very massive stars.
This trend is clearly evident in the stellar mass function evolution shown in Figure~\ref{fig:MF_2.35}, where smaller $\beta$ produces a more top-heavy distribution. 
The peak abundance values vary by species: carbon, oxygen, magnesium, silicon, sulfur are $\sim5-6$ times solar, nitrogen is $\sim4-20$ times solar, neon is $\sim10$ times solar, calcium is $\sim4$ times solar, and iron is $\sim3$ times solar. These enriched conditions persist for up to $\sim10^8\,$yr before declining. 
Stars formed during the overabundant phase may inherit unusually high and atypical abundance ratios, potentially giving rise to chemically distinct subpopulations.
Moreover, the physical origin of the nitrogen enhancement is further discussed in Section~\ref{sec:Nitrogen}, where we explore the implications of AMS enrichment for high-redshift galaxies.

Helium also exhibits a notable enhancement in AMS models. As shown in Figure~\ref{fig:He_2.35}, at high-$\dotmac$, [He/H] can reach 0.3-0.45 dex, corresponding to a helium mass fraction of 0.37 to 0.47. The peak amplitude and timing are both sensitive to the accretion parameter $\beta$, with lower values producing higher and earlier peaks, reflecting the enhanced early formation of massive stars under strong accretion conditions.

\subsection{Comparison with Conventional Chemical Evolution} \label{subsec:comparison}

\begin{figure*}
    \includegraphics[width=\textwidth]{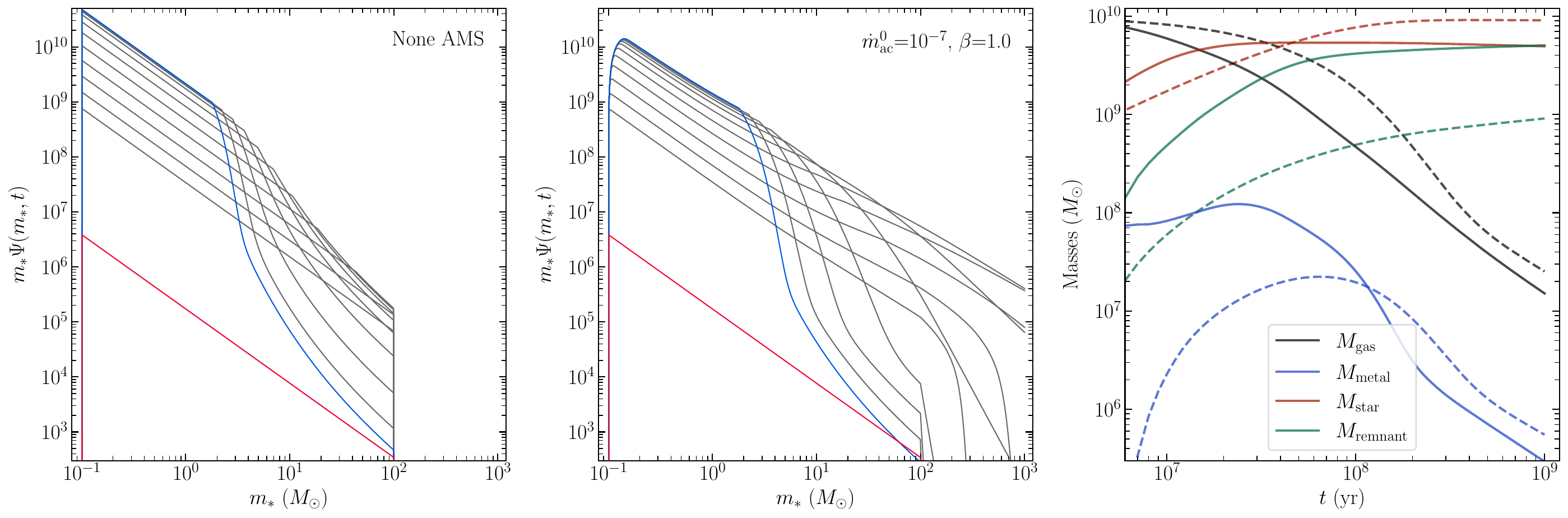}
    \caption{\footnotesize
    Comparison of conventional and AMS chemical evolution for a bulge with a total mass of $10^{10}\,\Msolar$, an SFE of 20 $\Gyr$, and a Salpeter IMF. Left panel: the evolving mass functions for conventional GCE. Middle panel: the evolving mass functions for AMS with $(\dotmac,\,\beta)=(10^{-7},\,1.0)$.
    Right panel: the time-dependent evolution of total gas mass, total gas-phase metal mass, total stellar mass, and total remnant mass plotted in black, blue, red, and green, respectively. The solid and dashed curves represent the AMS and conventional evolutions, respectively.}
    \label{fig:AMS}
\end{figure*}

One of the most distinctive differences between AMS and conventional chemical evolution is the emergence of a transient metallicity peak followed by a sharp decline (see Section~\ref{subsec:Metallicity}), which is absent in standard GCE scenarios and is analyzed in detail below.
In conventional chemical evolution, the gradual return of hydrogen from dying massive stars is accompanied by steady metal production and declining gas mass, resulting in a monotonic increase in metallicity.
In contrast, the AMS model introduces a burst of very massive stars that release large amounts of hydrogen into the ISM within a short timescale.
As accretion weakens and the formation of such stars quickly ceases, metal production drops significantly, while the sudden hydrogen return temporarily boosts the gas mass.
The resulting imbalance briefly reverses the metallicity growth and leads to a sharp decline before the system stabilizes again.

The comparison of mass function evolution between conventional and AMS chemical evolution is depicted in Figure~\ref{fig:AMS} (left and middle panels).
Here we adopt the AMS parameter set with the strongest stellar accretion effect, namely $(\dot{m}_{\rm ac}^0, \beta) = (10^{-7}, 1.0)$. After the transition from AMS mode to the conventional mode, the mass function distribution of the AMS model closely resembles that of the conventional model, indicating that it is observationally challenging to distinguish the presence of AMSs within bulges based solely on the mass function of existing stars. Extremely metal-rich low-mass stars formed during the early enrichment phase of AMSs may survive to the present day and could potentially be identified in the local universe (see Figure~\ref{fig:MF_2.35}). 
However, due to the significant uncertainties in stellar age determinations, it is also challenging to ascertain the presence of AMS in bulges by concurrently measuring the metallicity and age of stars.
This highlights the importance of combining mass function and chemical diagnostics when searching for observational signatures of AMS enrichment.

To further explore the implications of AMS enrichment, we show the evolution of total gas mass, gas-phase metal mass, stellar mass, and remnant mass for both conventional and AMS scenarios in the right panel of Figure~\ref{fig:AMS}. For the conventional evolution, gas is gradually converted into stars on a timescale of $10^8$ yr, consistent with theoretical predictions by \citet{Elmegreen1999}, and the gas-phase metal mass peaks at $\sim 2 \times 10^7\,M_\odot$, roughly matching the solar abundance benchmark from \citet{Ballero2008}. In comparison, the AMS model produces up to five times more metals compared to the conventional model.
In the late phases, both conventional and AMS scenarios reach the similar final gas mass $\sim 10^7\,M_\odot$, comparable to narrow-line region (NLR) gas masses ($\sim 10^{4-8}\,M_\odot$). Notably, AMS evolution results in a stellar mass roughly halved, but a remnant mass four times higher, implying that a substantial population of neutron stars and black holes may have formed in the early universe.
Observationally, evidence for a dense cusp of stellar-mass black holes has been found in the Galactic Center \citep{Hailey2018}, supporting the idea that compact remnants can efficiently accumulate in galactic nuclei.

\begin{figure*}
    \includegraphics[width=\textwidth]{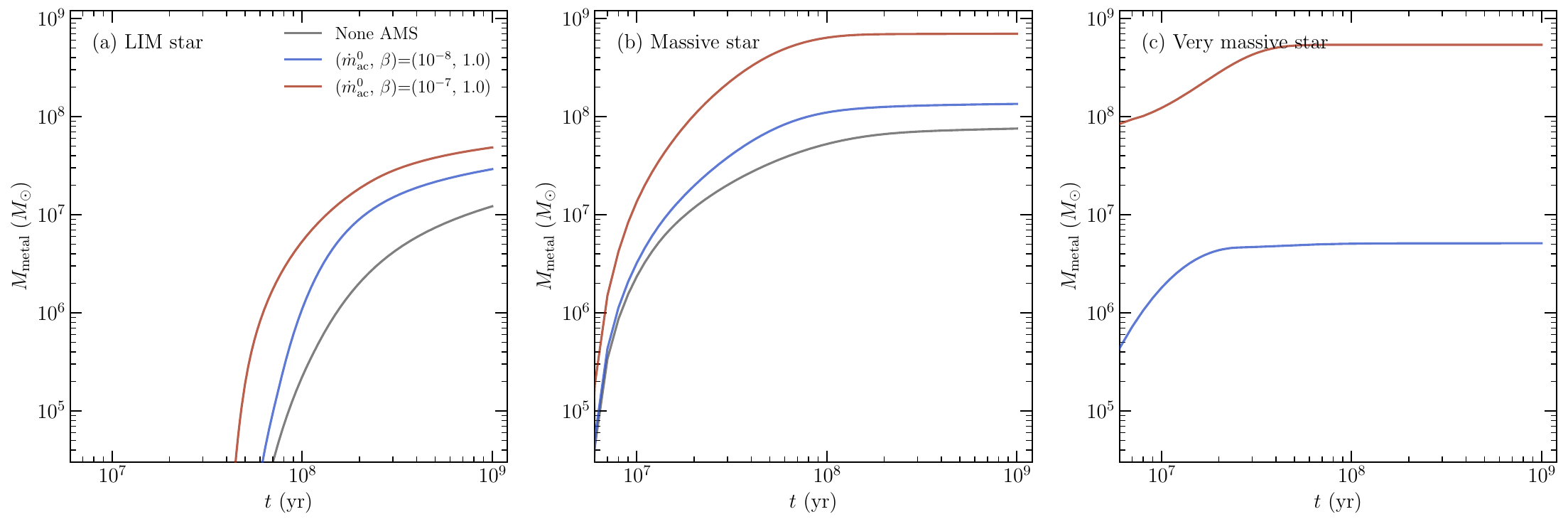}
    \caption{\footnotesize
    Time-dependence of metal mass contributed by stars of different masses for a bulge with a total mass of $10^{10}\,\Msolar$, an SFE of 20 $\Gyr$, and a Salpeter ($\alpha$=2.35) IMF. Gray, blue, and red curves denote the conventional GCE scenario, AMS with low-$\dotmac$ and AMS with high-$\dotmac$, respectively. Panel (a), (b) and (c) show the contributions from LIM stars ($m_* < 8\,M_\odot$), massive stars ($m_* \sim 10-100\,M_\odot$), and very massive stars ($m_* > 100\,M_\odot$), respectively.
    }
    \label{fig:metal_star}
\end{figure*}

To clarify the physical origin of the enhanced metal enrichment shown in Figure~\ref{fig:AMS}, Figure~\ref{fig:metal_star} further divides the total metal mass into contributions from stars of different masses. The results show that very massive AMSs ($m_{\ast}>100\,M_{\odot}$) dominate the early enrichment phase ($t\lesssim10^{7}\,\mathrm{yr}$), while massive and LIM stars contribute at later stages. This confirms that the super-solar peak in high-$\dot{m}_{\mathrm{ac}}^{0}$ models arises from the rapid feedback of very massive AMSs during the early evolution.
In addition, Figure~\ref{fig:metal_star} indicates that AMS evolution moderately enhances metal production across all stellar mass ranges compared with the conventional GCE model.

\section{Chemical Evolutionary Tracks in BPT Diagrams} \label{sec:VOresult}

\begin{figure*}
	\centering
	\begin{minipage}{0.95\linewidth}
		\centering
		\includegraphics[width=0.95\linewidth]{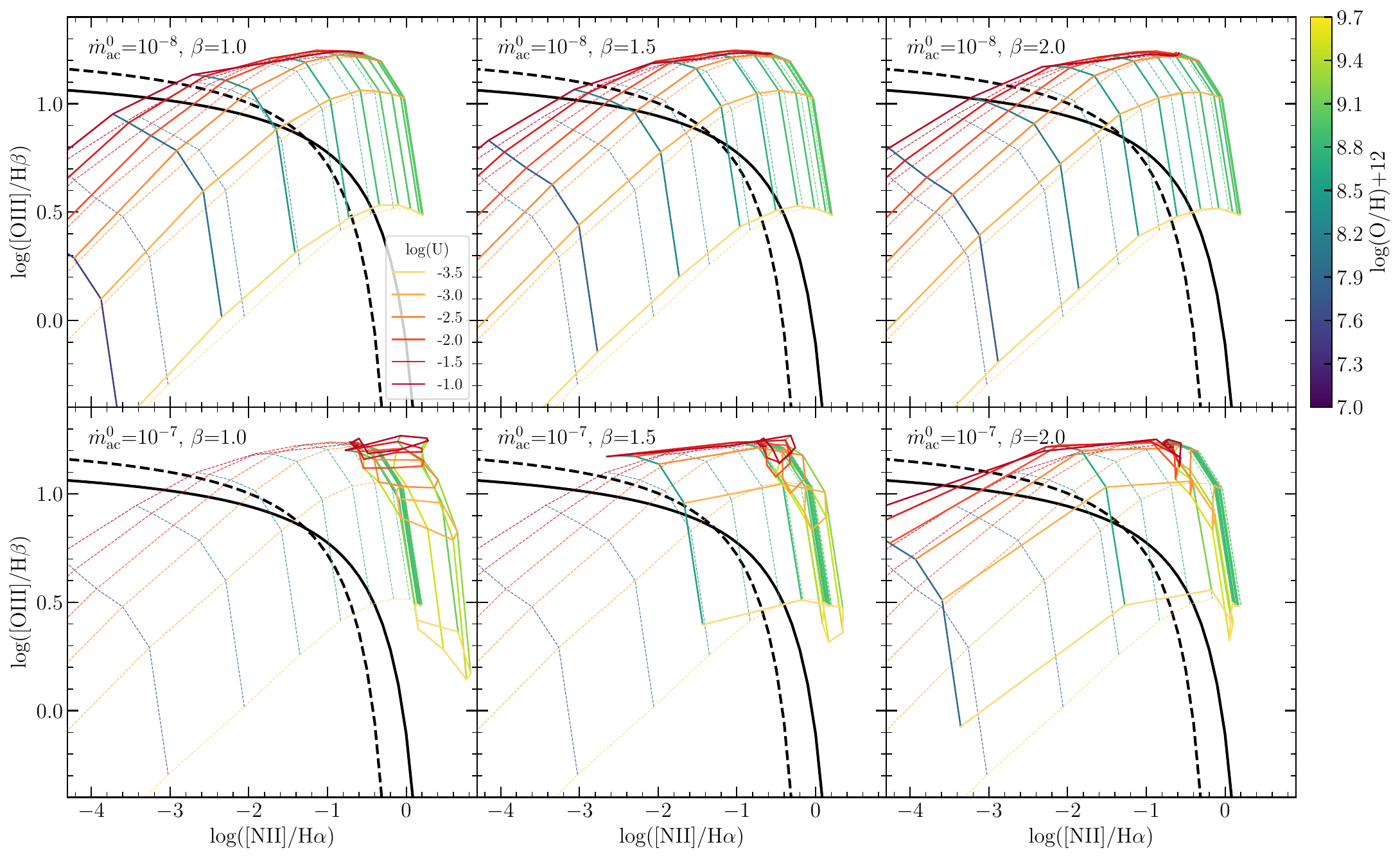}
		\caption{\footnotesize
                    The BPT diagrams with the colorbar representing oxygen abundance. Curves from yellow to red correspond to increasing ionization parameter, with log(U) varying from $-3.5$ to $-1.0$. The panels are arranged as follows: horizontal rows from left to right with increases of $\beta$ for a given $\dotmac$, and vertical with increases of $\dotmac$ from upper to below for a given $\beta$. The black solid curves are the theoretical maximum starburst lines proposed by \citet{Kewley2001}. The black dashed curve in the left column is the empirical maximum starburst line proposed by \citet{Kauffmann2003}. The black dashed lines in the right two columns are the Seyfert-LINER separation lines presented by \citet{Kewley2006}.}
		\label{fig:BPT_NIIHa}
        \vspace{5mm}
	\end{minipage}
	\begin{minipage}{0.95\linewidth}
		\centering
		\includegraphics[width=0.95\linewidth]{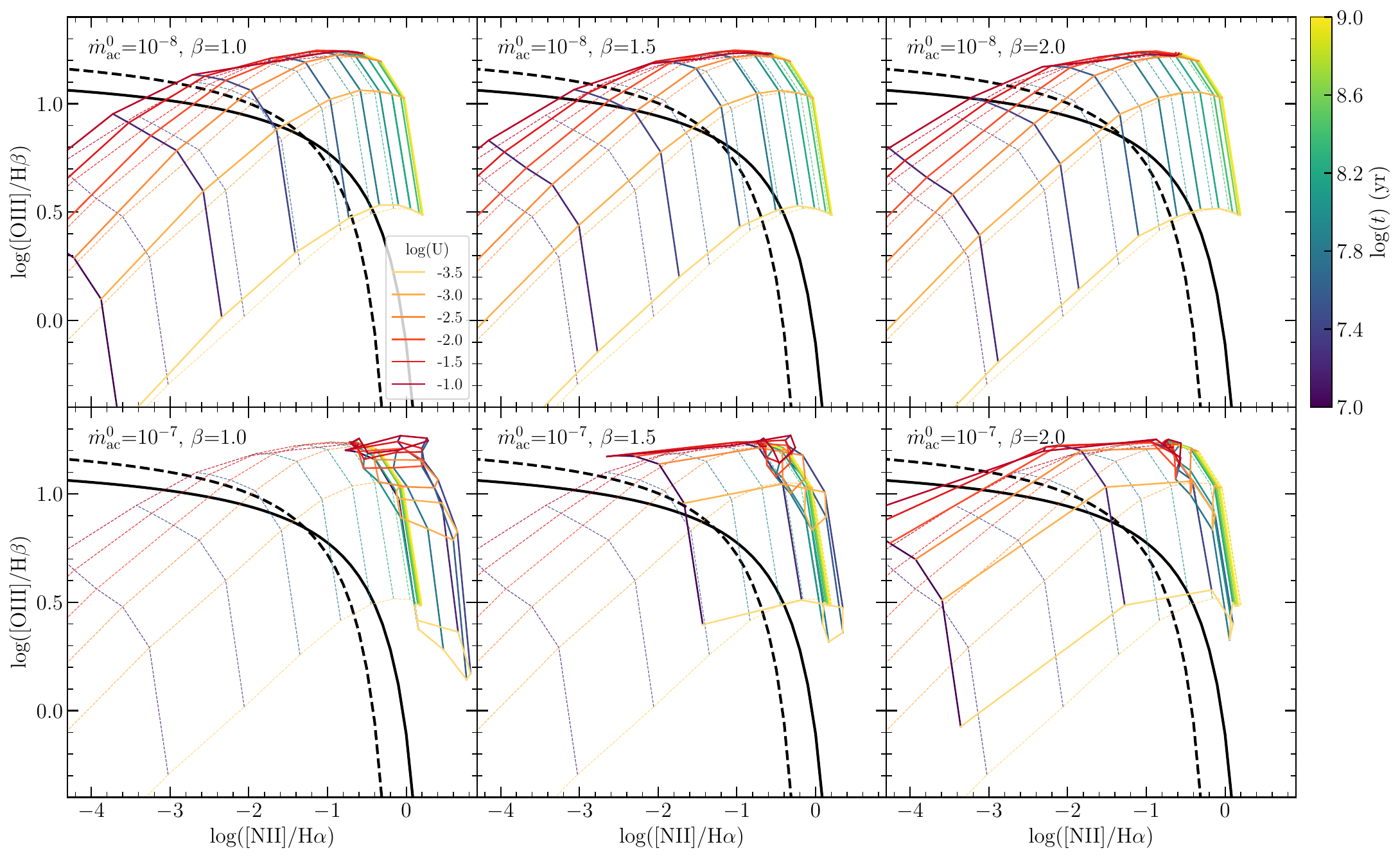}
		\caption{\footnotesize
                    The BPT diagrams with the colorbar representing time. Similar to Figure \ref{fig:BPT_NIIHa}.}
		\label{fig:BPT_NIIHa_time}
	\end{minipage}
\end{figure*}

In this section, we compare the chemical enrichment tracks of (i) conventional models versus those including AMSs, and (ii) AMS models with different parameter combinations. The goal is to explore whether AMS signatures can be identified in high-redshift AGN bulges using BPT diagrams.
Figures~\ref{fig:BPT_NIIHa}-\ref{fig:BPT_NIIHa_time} present the comparisons from two complementary perspectives, with panel arrangements increasing $\beta$ from left to right and increasing $\dotmac$ from top to bottom.
Figure~\ref{fig:BPT_NIIHa} uses oxygen abundance as colorbar to highlight differences in enrichment speed and maximum metallicity, while figure~\ref{fig:BPT_NIIHa_time} adopts evolutionary time as colorbar to emphasize the temporal evolution of the tracks.

Only at high-$\dotmac$ do the chemical evolution tracks differ significantly between AMS and conventional models, as shown in Figures~\ref{fig:BPT_NIIHa}-\ref{fig:BPT_NIIHa_time}.
Under such conditions, the BPT trajectories begin to deviate from the star-forming sequence after $\sim 10^7$–$10^8$ yr, shifting toward the local AGN branch \citep{Kewley2006} due to the rapid metal enrichment driven by massive stars, which is consistent with the metallicity evolution shown in Figure~\ref{fig:Z_AMS_IMF_nu}.
At high-$\dotmac$, $\beta$ also significantly affects both the early growth and the maximum value of [\ion{N}{2}]/H$\alpha$.
At $\beta=1.0$, log([\ion{N}{2}]/H$\alpha$) can reach 0.8\,dex, $\sim 0.6\,$dex higher than in conventional evolution, while the maximum value is the same as in conventional evolution at $\beta=2.0$. 
Although early-time line ratios differ markedly, all models converge in metallicity by 1\,Gyr, restricting the detectability of AMS signatures to brief time intervals.

A detailed inspection of Figures~\ref{fig:BPT_NIIHa}-\ref{fig:BPT_NIIHa_time} reveals that the parameter combination $(\dotmac,\,\beta)=(10^{-7},\,1.0)$ yields the clearest observational signature of AMSs. 
A dense gas environment is required to sustain stellar accretion and maintain high gas-phase metallicity for approximately $10^8$ yr (see Figure~\ref{fig:Z_AMS_IMF_nu}), increasing the likelihood of detecting AMS signatures at high redshifts ($z \gtrsim 15$). This parameter set therefore offers a promising spectroscopic pathway to probe AMS populations in AGN host bulges.

The large offsets in [\ion{N}{2}]/H$\alpha$ cannot be attributed to variations in the AGN radiation field alone, as demonstrated in studies of ionization parameter and Eddington ratio \citep{Thomas2016}. 
The chemical enrichment history thus plays a dominant role in shaping BPT trajectories in the early universe.

\section{Discussion} \label{sec:modeldependences}

To clarify the assumptions and limitations of our current modeling framework, we discuss five physical processes that are not included in our closed-box chemical evolution model but could potentially impact the evolution of AGN host bulges: the infall of primordial or low-metallicity gas, feedback from supernovae, AGN feedback, contribution from binary stars, and uncertainties in stellar nucleosynthetic yields.
Although beyond the scope of this study, each process offers a promising direction for future modeling and interpretation of observational data.
In addition, we highlight the issue of nitrogen enrichment, which has emerged as a key observational feature of high-redshift galaxies and provides an important benchmark for testing AMS-driven chemical evolution.

\subsection{Infall} \label{subsec:infall}

The accretion of galaxies is considered to be an important pathway in the formation of bulges and an effective way to dilute gas-phase metals {\cred \citep{Dekel2009Natur}}. In GCE models, the infall of the extragalactic medium into the gravitational potential of a galaxy is commonly described by a decaying-time exponential-like model \citep{Matteucci2021,Spitoni2021}. 
Gas infall is commonly associated with triggering new star formation; however, the accreted primordial or low-metallicity gas may also be partially consumed by existing stars through stellar-scale accretion, potentially forming AMSs.
If the infalling gas is sufficiently dense and abundant, this process can accelerate the recovery of gas-phase metallicity from its initially diluted state, and in some cases, even result in an overall metallicity increase within the bulge. 
Achieving this outcome depends on the efficiency of coupling between the gas and the stellar population in the bulge, which may vary across systems.

Galaxies may be formed by one or more separated accretion episodes, typically modeled by decaying-time exponential-like infalls of gas. With the advent of several large spectroscopic surveys, such as LAMOST \citep{Zhao2012}, GALAH \citep{DeSilva2015}, SDSS/APOGEE \citep{Majewski2017}, and Gaia DR3 \citep{RecioBlanco2023}, it is becoming increasingly evident that there are two sequences in the [$\alpha$/Fe]-[Fe/H] with systematic differences in stellar ages, indicative of multiple gas accretion events in the Milky Way \citep{Noguchi2018,Spitoni2019,Spitoni2023,Lian2020a,Lian2020b}. Furthermore, JWST spectroscopy has unveiled a population of AGN at $z\textgreater4$ with sub-solar NLRs \citep{Harikane2023, Maiolino2023, Marshall2023, Ubler2023, Kocevski2023}, suggesting the accretion of pristine or metal-poor gas at high redshift. Infall events may therefore play an important role in fueling AMS formation and regulating the metal enrichment history of host bulges, particularly in the early universe. However, the incorporation of gas infall is beyond the scope of this work. Future extensions will explicitly account for the infall to provide a more complete picture of the evolution and enrichment of AMS in bulges.


\subsection{Feedback from Supernovae} \label{subsec:SNefeedback}

Supernova-driven galactic winds are typically considered in GCE models to suppress star formation, i.e., star formation stops when cumulative thermal energy injected by supernovae exceeds the binding energy of the galactic gas \citep{Matteucci1987}.
Galactic winds typically onset $\sim 10^8-10^9\,$yr after galaxy formation, after which gas-phase metallicities cease to increase and may even decline due to dilution from stellar mass loss.
However, whether supernova feedback is involved in bulge chemical evolution remains uncertain.
\citet{Elmegreen1999} sustained that the potential well of the bulge is too deep for self-regulation, suggesting the inclusion of galactic winds is not necessary. 
\citet{Ballero2007} included Supernovae feedback in their calculations but found that galactic winds occur when the majority of gas has already been converted into stars, rendering their effect negligible.
We therefore neglect galactic winds in our model.

Recently, \citet{Li2024} proposed a feedback-free starburst scenario to explain the observed excess of bright galaxies at $z \gtrsim 10$ in JWST observations, inconsistent with standard feedback-regulated models. 
Therefore, excluding supernova feedback is not merely a simplifying assumption, but a physically motivated choice consistent with current high-redshift observations. This further supports our exclusion of supernova feedback in the present model.


\subsection{AGN Feedback}

AGN feedback can influence the evolution of bulges by either quenching or triggering star formation \citep{Ciotti2007}. Theoretical calculations and hydrodynamical simulations regarding AGN feedback are abundant \citep{Silk2012, Li2018, Valentini2021, Ciotti2017, Ciotti2022}.

In studies of chemical evolution, the inclusion of AGN feedback depends on phenomenological treatment, considering only radiative feedback and neglecting other feedback mechanisms such as radiation pressure and relativistic particles, as well as the phenomena associated with jets \citep{Padovani1993, Ballero2008, Molero2023}. 
Studies indicate that AGN feedback has a negligible effect on chemical evolution. 
Even in the few sources showing evidence of AGN feedback, the overall impact does not appear to globally suppress star formation \citep{Cresci2015}. 

For high-redshift AGNs, AGN-driven outflows are revealed by \citet{Maiolino2023arXiv} and \citet{Matthee2023}. For instance, \citet{Ubler2023} detected an outflow with projected velocity $\gtrsim 700 \, \mathrm{km/s}$ in galaxy GS\_3073 and pointed out this outflow may suppress fresh gas accretion but contributes little to global feedback. 
Therefore, neglecting AGN feedback in our model does not compromise the validity of our chemical evolution predictions.

\subsection{Binary System}

Binary systems are commonly included in chemical evolution models, as Type Ia supernovae are among the primary sources of iron. However, in the context of high-redshift ($z \gtrsim 10$) bulge evolution, their contribution is expected to be subdominant. Type Ia supernovae exhibit delayed time distributions, typically becoming significant only $0.1–1$ Gyr after star formation \citep{Matteucci1986,Mannucci2006}, while the early evolutionary phases modeled here occur before such timescales elapse. In addition, observational constraints on the binary fraction, delay-time distribution, and associated feedback mechanisms in the high-redshift universe remain uncertain \citep{Maoz2014}. Therefore, neglecting the contribution of binary systems represents a simplified yet physically justified assumption in our model.

\subsection{Stellar Nucleosynthetic Yields}

As nucleosynthetic yields for AMSs have not yet been computed, we adopt the yields of normal stellar populations as a proxy in this work. Several recent simulations suggest that AMSs (i.e., stars embedded in AGN accretion disks) can undergo rapid evolution and produce significant amounts of heavy elements \citep{Cantiello2021,Fryer2025}, which may significantly alter the chemical enrichment pathways of their host environments. However, these studies primarily focus on the stellar structure and evolution, without explicitly calculating their nucleosynthetic outputs. As such, the chemical evolution results presented here should be regarded as a first-order approximation. Future studies are needed to quantify how the yields of AMSs may differ from those of normal stars and how such differences could affect the resulting abundance patterns.

\subsection{Nitrogen Enrichment at High Redshift: Clues and Implications} \label{sec:Nitrogen}

As described in Section~\ref{subsec:AMSAbun}, AMS models with higher accretion rates exhibit enhanced [N/H] and [N/O] ratios. Here we further explore the physical origin and implications of this nitrogen enrichment.

Recent JWST spectroscopy has revealed widespread nitrogen enhancement in high-redshift galaxies. Enhanced nitrogen-to-oxygen (N/O) has been reported in multiple early galaxies \citep{Bunker2023, Cameron2023, Tacchella2023, Hayes2025, Ji2025, Roberts2025}. This observational evidence highlights a growing challenge for models of early chemical evolution.

The origin of nitrogen enrichment remains debated, with proposed contributions from CCSNe, intermediate-mass asymptotic giant branch stars, and rotating massive stars \citep{Chiappini2006, Nomoto2013}. Recently, the enhanced N/O galaxies uncovered by JWST have renewed interest in scenarios invoking very massive stars and/or a variable initial mass function, which naturally yield elevated nitrogen abundances and have been proposed to explain the observations \citep{Charbonnel2023, Nandal2025}.


Our AMS models at high-$\dotmac$ show nitrogen enhancement during the early evolutionary stages, although they do not reach the extreme N/O ratios observed in some high-redshift galaxies (see Figure~\ref{fig:Metal_2.35}). The discrepancy in N/O mainly reflects the limitations of the adopted stellar evolution and nucleosynthetic yield prescriptions. Nevertheless, AMS enrichment naturally boosts nitrogen production, consistent with the direction suggested by very massive star scenarios, which indicates that AMSs may represent a complementary channel to the nitrogen problem in high-redshift galaxies once more advanced stellar yields are incorporated into the models.

\section{Conclusions} \label{sec:concl}

We develop a new chemical evolution model that incorporates AMSs, a theoretically motivated stellar population embedded in AGN accretion disks. By tracking the gas-phase metallicity evolution in a closed-box framework over $1\,$Gyr and coupling it with photoionization modeling, we use the BPT diagram to predict potential observational signatures of AMSs in AGN host bulges. These predictions may serve as theoretical guidance for future efforts to identify AMSs at high redshift. Our main findings are summarized as follows:
\begin{itemize}
  \item[1.]AMS mass functions in the bulge do not appear as top-heavy distributions since the relatively low density of the bulge results in a lower AMS accretion rate. As the gas density decreases over time in the closed-box model, the AMS accretion rate declines accordingly, making it increasingly difficult for top-heavy mass functions to emerge. At later evolutionary stages, the system transitions to the conventional mode, and the resulting AMS mass functions become indistinguishable from those of normal stars. This suggests that stellar mass functions alone may not provide a viable way to test the presence of AMSs.
  \item[2.]AMSs allow the gas-phase metallicities of host bulges to reach solar levels more rapidly. Particularly under high accretion conditions, they can generate metallicity bumps in the early stages of evolution, with element abundances reaching 4–10 times solar values and persisting for approximately 0.1\,Gyr. In addition, the abundance ratios of various elements undergo significant changes as AMS accretion rates increase, potentially resulting in the formation of stars with atypical abundance ratios.
  \item[3.]Compared to BPT diagrams using conventional chemical evolution, AMS at high accretion rates can cause the coverage to shift entirely to the local AGN branch after $10^7-10^8\,$yr, and significantly extend the upper limit of the high metallicity region in the BPT diagram, with AGNs at $z\gtrsim15$ expected to occupy locations overlapping those of local AGNs.
  This suggests that the detection of super-solar NLRs in extremely high-redshift AGNs ($z\gtrsim15$) may provide tentative evidence for the existence of AMS in their host bulges.
\end{itemize}

Although predicted chemical features remain observationally challenging, the relevant wavelengths are already covered by JWST/MIRI \citep{Wright2023}, offering an opportunity to search for the most prominent lines in exceptionally luminous sources, despite the limited sensitivity near $\sim 10 \mu m$. Next-generation mid-infrared telescopes, ELT/METIS \citep{Brandl2021} and the cryogenically cooled Origins Space Telescope \citep{Battersby2018}, are expected to deliver orders-of-magnitude sensitivity gains, thereby enabling systematic studies of these spectral signatures at very high redshifts. The predicted chemical signatures thus provide concrete and timely targets for forthcoming spectroscopic surveys and motivate continued theoretical development and coordinated observational efforts to assess the role of AMSs in the early chemical enrichment of AGN bulges.

\section{Acknowledgments}


S.Z. is grateful to P.-X. Zhu, W.-Y. Xin, J.-Z. Zhou, X.-X. Xue and H.-N. Li for valuable discussions.
J.M.W. thanks B-type Strategic Priority Program of the Chinese Academy of Sciences (Grant No. XDB1160202).
This study is supported by the National Natural Science Foundation of China under grant Nos. 12588202, 12333003, and the National Key R\&D Program of China
under grant Nos. 2024YFA1611900, 2021YFA1600404.

\appendix

\section{Stellar Lifetime} \label{app:lifetime}

\begin{figure*}
\centering
    \includegraphics[width=0.5\textwidth]{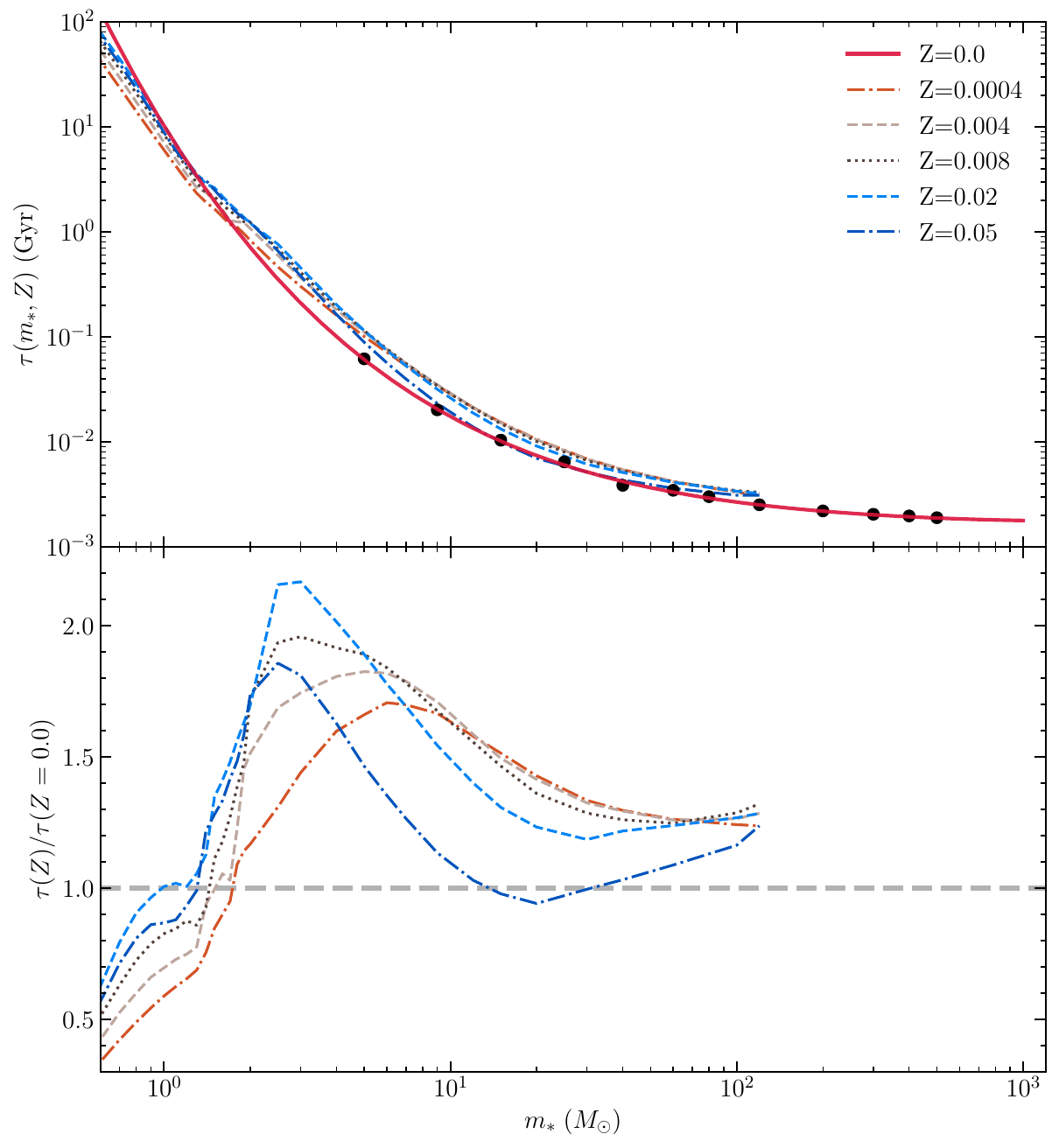}
    \caption{\footnotesize
    Top panel: metallicity-dependent stellar lifetime $\tau (m_*,Z)$ from \citet{Schaerer2002} (red line, $Z=0$) and \citet{Portinari1998} ($Z=0.0004, 0.004, 0.008, 0.02, 0.05$). Bottom panel: the ratio of the five metallicity-dependent stellar lifetimes from \citet{Portinari1998} normalized by the one from \citet{Schaerer2002}.}
    \label{fig:stellarlifetime}
\end{figure*}

We employ the theoretical results of main-sequence lifetimes of Population III stars from \citet{Schaerer2002}. 
To test the impact of metallicity on the stellar lifetime, we compare the main-sequence lifetime$-$mass relation of Population III from \citet{Schaerer2002} and those of Population I and II from \citet{Portinari1998}. 
As demonstrated in Figure \ref{fig:stellarlifetime}, the polynomial fitting result is approximately consistent with the main-sequence lifetime$-$mass relation from \citet{Portinari1998} for $\ms<120\,\Msolar$, fluctuating between 0.5-2 times. 
The lifetime of very massive stars is typically several Myr. 
These results suggest that main-sequence lifetimes are nearly metallicity-independent, validating the use of the Population III lifetime–mass relation.

\section{Abundance Test} \label{app:testAbn}

\begin{figure*}
    \centering
    \includegraphics[width=0.9\textwidth]{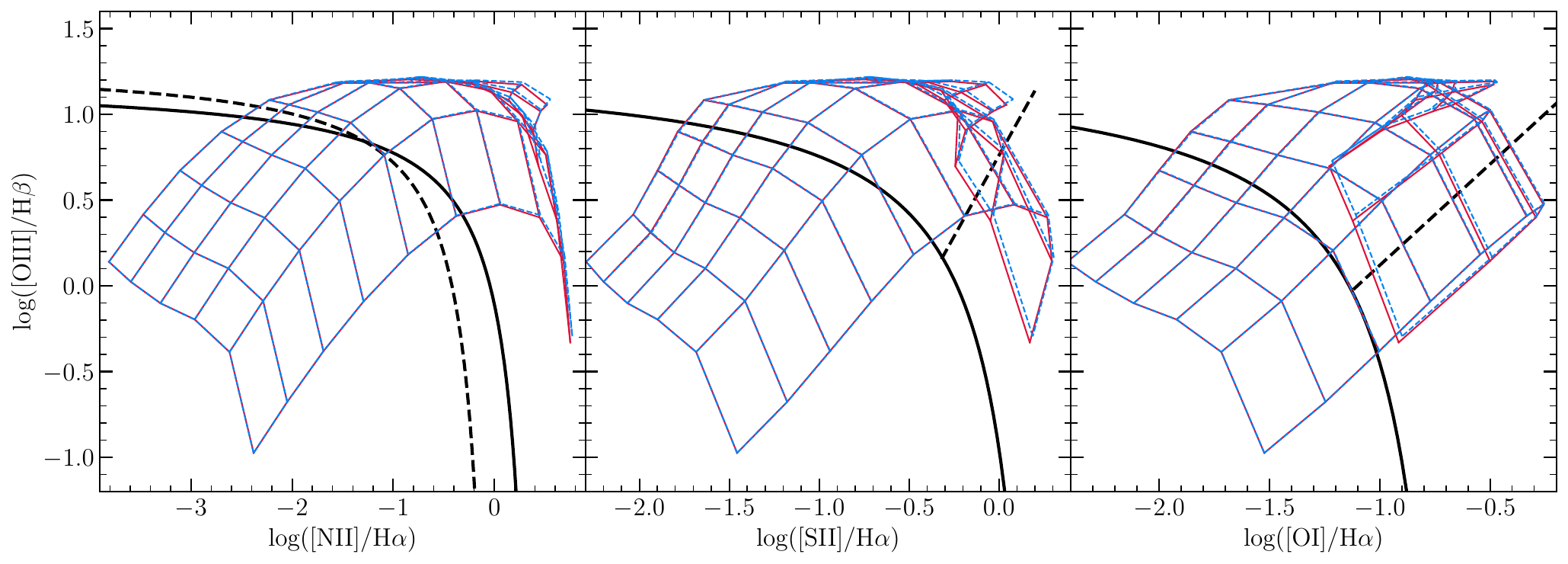}
    \caption{\footnotesize
    Comparison of AGN models built with the nonlinear scaling relation of \citet{Nicholls2017} (solid red) and that lacking elements (dashed blue) on the standard optical diagnostic diagrams. Each model is presented with constant-metallicity lines ($\text{12+log(O/H)}=7.0,\, 7.3,\, 7.6,\, 7.9,\, 8.2,\, 8.5,\, 8.8,\, 9.1,\, 9.4,\, 9.7$) and constant-ionization-parameter lines ($\text{log(U)}=-3.5,\, -3.0,\, -2.5,\, -2.0,\, -1.5,\, -1.0$). The black solid curves are the theoretical maximum starburst lines proposed by \citet{Kewley2001}. The black dashed curve in the left panel is the empirical maximum starburst line proposed by \citet{Kauffmann2003}. The black dashed lines in the right two panels are the Seyfert-LINER separation lines presented by \citet{Kewley2006}.}
    \label{fig:BPT_testAbn}
\end{figure*}

Among the 30 elements included in the MAPPINGS database, we do not compute the abundances of Li, Be, B, F, Na, Al, P, Cl, Ar, K, Sc, Ti, V, Cr, Mn, Co, Ni, Cu, and Zn. To assess whether these neglected elements impact the optical diagnostic diagrams, we adopt the AGN radiation field with a high Eddington ratio and the nonlinear scaling from \citet{Nicholls2017}. We then compare two sets of models: one with standard abundances and one with the above elements set to $\log_{10}(X/H) = -10$, effectively rendering them negligible. The remaining settings follow those in Section~\ref{sec:Photoionization}.
As shown in Figure~\ref{fig:BPT_testAbn}, the exclusion of these elements slightly increases [\ion{O}{3}]/H$\beta$ at high metallicity, consistent with a modest rise in the ionization parameter due to lower overall metal content. We conclude that the omission of these elements introduces minimal bias, validating the metallicity assumptions adopted in Section~\ref{sec:Photoionization}.

%

\bibliography{ref}{}
\bibliographystyle{aasjournal}
\end{document}